\documentclass[journal]{IEEEtran}

\usepackage{cite}

\usepackage[pdftex]{graphicx}

\usepackage[T1]{fontenc}
\usepackage[cmintegrals]{newtxmath}
\usepackage{amsmath}
\usepackage{citesort}
\usepackage{ulem}
\usepackage{amsfonts}

\usepackage{amssymb,bm,upgreek}
\usepackage{subfigure}
\usepackage{multirow}
\usepackage{array}
\usepackage{xcolor}
\usepackage{latexsym}
\usepackage{tabularx}
\usepackage{stfloats}
\usepackage{epic}
\usepackage{dashrule}	
\usepackage{gensymb}
\usepackage{booktabs}
\usepackage{url}
\usepackage{makecell}

\begin{document}

\title{A Tri-Band Shared-Aperture Base Station Antenna Array Covering 5G Mid-Band and 6G Centimetric Wave Band}

\author {Shang-Yi~Sun,~\IEEEmembership{Graduate~Student~Member,~IEEE},
        Hai-Han~Sun,~\IEEEmembership{Senior Member,~IEEE},
        Can~Ding,~\IEEEmembership{Senior Member,~IEEE},        
        and~Y.~Jay~Guo,~\IEEEmembership{Life Fellow,~IEEE}

\thanks{This work was supported by the Australian Research Council (ARC) Discovery Early Career Research Fellowship (DECRA) under
Grant DE200101347. \emph{(Corresponding author: Can Ding.)}}

\thanks{Shang-Yi Sun, Can Ding, and Y. Jay Guo are with the Global Big Data Technologies Centre, University of Technology Sydney, NSW 2007, Australia (e-mail: Shangyi.Sun@student.uts.edu.au; can.ding.1989@gmail.com; Jay.Guo@uts.edu.au).}

\thanks{Hai-Han Sun is with the Department of Electrical and Computer Engineering, University of Wisconsin-Madison, Madison, WI 53706, USA (e-mail: Haihan.Sun@wisc.edu).}

}

%

\maketitle

\begin{abstract}
This work proposes a tri-band shared-aperture antenna array with three wide bands, covering the 5G mid-band and the 6G centimetric band, which is a promising candidate for future 6G base station antennas. The challenge of suppressing interferences, including scattering and coupling, in the tri-band array is holistically addressed across wide bands. Guided by characteristic mode analysis (CMA), a segmented spiral radiator is efficiently developed to mitigate scattering and coupling at high frequencies while preserving radiation performance at low frequencies. Compared to a conventional tube radiator, the proposed spiral exhibits a reduced radar cross-section (RCS) over an ultra-wide range of 4.7-21.5 GHz (128.2\%). With the aid of serial resonators, impedance matching of the segmented-spiral-based dipole antenna is achieved across the low band (LB) of 3.05-4.68 GHz (42.2\%), spanning the 5G band 3.3-4.2 GHz. Moreover, suppressors are placed near the LB ports to further reduce the cross-band coupling. Middle band (MB) and high band (HB) antennas operate in 6.2-10.0 GHz (46.9\%) and 10.0-15.6 GHz (43.8\%), respectively, collectively covering the anticipated 5G-Advanced and 6G centimetric band of 6.425-15.35 GHz. Both the MB and HB antennas employ a planar magnetoelectric (ME) dipole structure, which prevents common-mode resonances in the LB and MB, and mitigates the scattering from the MB antenna in the HB. In this tri-band array, radiation patterns remain undistorted across the LB, MB, and HB, and the isolation between any two ports exceeds 20 dB over all three bands.
\end{abstract}


%
\IEEEpeerreviewmaketitle

\section{Introduction}

\IEEEPARstart{T}{he} rapid evolution of the mobile communication technology, coupled with the continuous pursuit of cost-efficiency and miniaturization, requires antennas operating at different frequencies to share an extremely limited space to simultaneously support various standards. The co-existence of different antennas results in cross-band interference, namely scattering and coupling, which leads to severe distortion of radiation patterns, and degradation of isolation and impedance matching. Wide-band suppression of the cross-band scattering and coupling remains a critical challenge in designing high-performance multi-band antenna arrays.

The cross-band scattering in the dual-band array occurs on both the low band (LB) and high band (HB) antennas. To suppress the HB scattering caused by LB antennas, cloaks \cite{S1}, \cite{S2}, \cite{S3}, slots \cite{S4}, \cite{S5}, L-shaped branches \cite{S6}, chokes \cite{S8}, \cite{S9} or frequency selective surfaces (FSS) \cite{S10}, \cite{S11} are co-designed with the LB antennas to generate reversed currents, obstruct induced currents or form passband in the HB. On the other hand, the LB scattering results from the common-mode resonance of the HB antennas in the target LB. Connecting an inductor \cite{S3} or a capacitor \cite{S12} in series with the HB balun can shift the resonance out of the LB. Moreover, the reduction of cross-band coupling is typically achieved through filtering techniques to generate radiation nulls or restraint coupling currents flowing into ports \cite{C2}, \cite{C3}, \cite{C4}. However, they are mainly designed to enhance isolation without contributing much to the scattering suppression.

To simultaneously suppress cross-band scattering and coupling in dual-band arrays, stacked configurations with the HB antenna placed above the LB antenna are proposed in \cite{A1}, \cite{A6}, \cite{A5} and \cite{A4}, smartly avoiding the cross-band interference issues. In addition, dual-functional structures have been designed without changing the conventional interleaved configuration. The folded-dipole in \cite{C3} and U-shaped structure in \cite{D1} serve as the LB antennas that allow HB wave propagation without interference and also prevent HB current from flowing into LB ports.

For the tri-band array, the suppression of cross-band scattering and coupling is much more complex than in dual-band arrays, as interferences occur between any two of the three types of antennas. Currently, works on tri-band arrays achieve the suppression by employing various methods proposed for dual-band arrays. To suppress scattering in both the middle band (MB) and HB, the FSS structures featuring dual passbands are integrated into the LB radiators \cite{T5}, \cite{T6}, \cite{T4}. In \cite{T1}, chokes, two groups of slots, and series inductors are adopted to reduce different forms of scattering, while slots and FSS-based MB radiators are utilized in \cite{T3}. The stacked configuration is adopted by \cite{T2} and \cite{T7} to avoid scattering and coupling problems. Due to the increased challenges, existing tri-band array works still suffer from relatively narrow bandwidths, and the problem of cross-band coupling has not been fully addressed.

6G has now entered the research stage, with expectations to introduce several frequency bands within 7.125-15.35 GHz to achieve a balance between capacity and coverage \cite{6G1}, \cite{6G3}, \cite{6G2}. Moreover, the band 6.425-7.125 GHz has been allocated mainly for the 5G-Advanced and may serve a foundational band for 6G in certain countries \cite{6G3}. The higher, wider and more numerous operating bands for 5G and 6G pose new challenges for the design of antenna elements and multi-band arrays. For instance, if antennas continue to adopt the commonly used 3D dipole, the significantly reduced size will introduce difficulties in fabrication and assembly. More importantly, the use of additional and wider bands further complicates the interference problems within the antenna array. 

\begin{figure}[!t]
\centering
\includegraphics[trim=0 10 0 0,width=8.8cm]{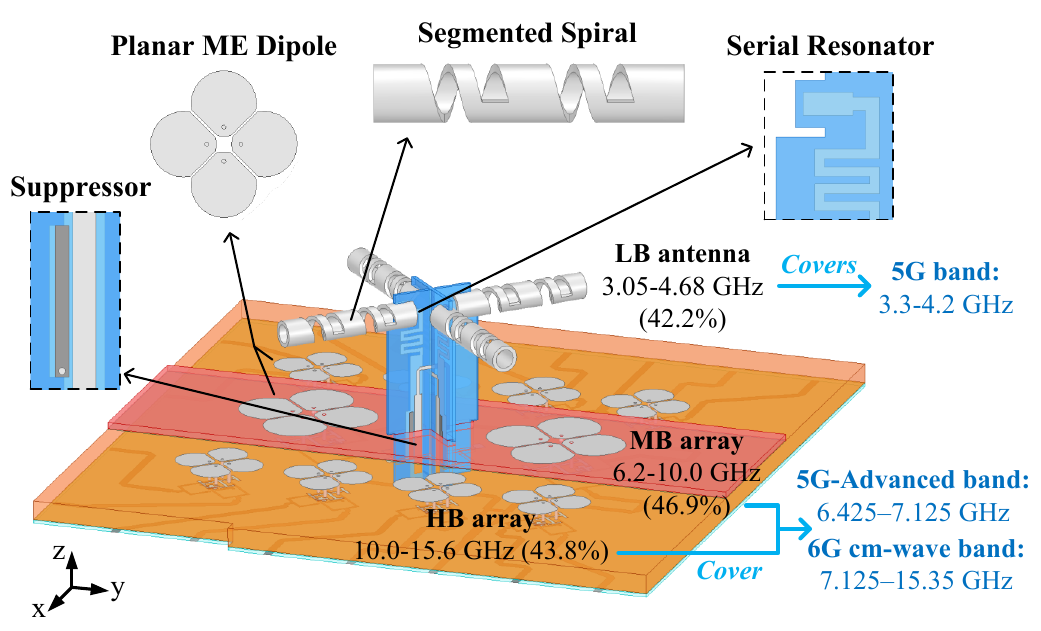}
\caption{Configuration of the developed tri-band antenna array.}
\label{Tri_Band_Array}
\end{figure}

In this paper, as shown in Fig. \ref{Tri_Band_Array}, a tri-band 5G/6G antenna array is developed with operation bandwidths in the LB, MB, and HB ranges of 3.05-4.68 GHz (42.2\%), 6.2-10.0 GHz (46.9\%), and 10.0-15.6 GHz (43.8\%), respectively. The radiation pattern distortions caused by scattering within the LB, MB, and HB have been effectively mitigated or avoided, and all cross-/in-band coupling has been suppressed to below -20 dB. The novelty and contributions are outlined as follows:

(i) \textbf{Ultra-Wide-Range in All Three Bands, Enabling Full Coverage for 5G and 6G:} The proposed tri-band shared-aperture antenna array features the widest bandwidths among existing designs. The LB antenna supports the 5G sub-6 GHz band 3.3-4.2 GHz, while the MB and HB antennas collectively cover the 5G-Advanced and 6G range of 6.425-15.35 GHz, ensuring seamless multi-generation connectivity. 

(ii) \textbf{Electromagnetic Transparency via Characteristic Mode Analysis (CMA):} To prevent the LB radiator from distorting the radiation of MB and HB antennas, the LB segmented spiral radiators are strategically designed using the CMA to exhibit electromagnetic transparency across both the MB and HB. This design achieves a radar cross-section (RCS) reduction across 4.7–21.5 GHz (128.2\%), demonstrating ultra-wideband scattering suppression.

(iii) \textbf{Impedance Compensation for Matching:} To address the degradation in LB impedance matching caused by the scattering suppression across an ultra-wide band, a pair of serial resonators within the LB balun is designed. These resonators effectively compensate for the impedance mismatch, achieving a matching bandwidth of 42.2\%. Additionally, suppressors are placed near the LB ports to further reduce cross-band coupling between the LB and MB antennas. 
 
(iv) \textbf{Low-Profile Planar ME Dipole Design:} Both the MB and HB antennas adopt a planar magnetoelectric (ME) dipole structure. This low-profile design, simplifying the assembly process, prevents common-mode resonances in both the LB and MB, and reduces the scattering from the MB antenna in the HB.

This paper is organized as follows. Section II and III detail the design of the LB and MB/HB antennas, respectively. The scattering and coupling suppression results in the tri-band array are provided in Section IV. Section V presents the experimental results of the fabricated antenna array. Finally, the conclusions are summarized in Section VI.

\section{LB Antenna Design}
\subsection{LB Radiator Design Guided by Characteristic Mode Analysis (CMA)}

When the LB antenna is positioned adjacent to the radiating MB/HB antennas, MB/HB currents are induced on the LB radiator. These induced currents give rise to secondary radiation, which interferes with the original MB/HB radiation and results in distorted total radiation patterns in the MB/HB. This phenomenon is known as cross-band scattering in the MB/HB. The induced currents on the LB radiator also flow into antenna ports through the LB balun, causing cross-band coupling in the MB/HB \cite{Book}. Furthermore, the MB/HB secondary radiation from the LB radiator also excites induced currents on the other MB/HB antennas, intensifying in-band coupling between MB/HB antennas. Consequently, the cross-band scattering, cross-band coupling, and in-band coupling in the MB/HB are all attributed to the MB/HB induced currents on the LB radiators.

According to the characteristic mode theory, the total induced currents on the LB radiator under the given MB/HB excitation can be approximated by combining currents of a few significant modes with larger modal weighting coefficient (MWC) values in the MB/HB \cite{CMA1}. The desired interference suppression can be accomplished by reducing the |MWC| values associated with the significant modes \cite{CMA2}, \cite{CMA3}. In the MWC simulation, a plane wave is used as the excitation source for the LB radiator, providing the most effective and simplest approximation of the radiated waves from the MB/HB antennas. It is worth noting that MWC is more suitable for addressing the scattering suppression problem compared to modal significance (MS), as the excitation source in this problem is predetermined. In contrast, MS considers all the modes of the structure itself, many of which cannot be excited by this specific excitation source. The inclusion of too many redundant modes complicates observation and analysis.

\begin{figure}[!t]
\centering
\includegraphics[trim=0 10 0 0,scale=1]{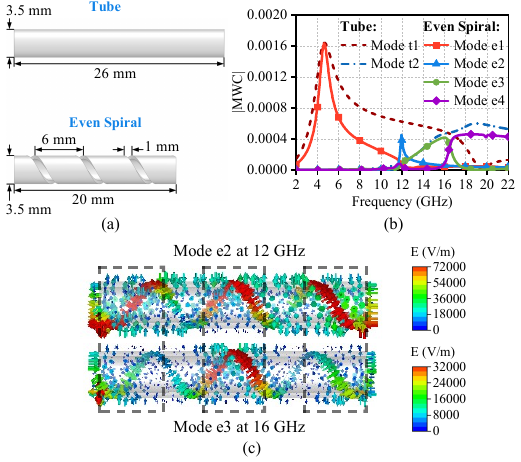}
\caption{(a) Geometry of the tube and even spiral, (b) |MWC| of the tube and even spiral, and (c) modal E-field distribution of Mode e2 and Mode e3.}
\label{CMA_Even}
\end{figure}

The cylindrical tube shown in Fig. \ref{CMA_Even}(a) is a traditional structure used as the radiating arm of the dipole antenna. It can be matched within the LB as antenna radiator, and it will be compared as a reference case in the following sections. The Mode t1 shown in Fig. \ref{CMA_Even}(b) is the dominant mode of the tube. Its value remains high at higher frequencies, suggesting that the tube can be easily excited by waves at higher frequencies. The target of this work is to modify the tube-based radiator to expand the scattering and coupling suppression band as much as possible while maintaining the matching performance of the LB antenna. Therefore, it is imperative to significantly reduce the |MWC| values of Mode t1 at higher frequencies while keeping the |MWC| values at lower frequencies unchanged.

As shown in Fig. \ref{CMA_Even}(a), the spiral structure previously investigated in \cite{S9} appears to offer a promising solution for the suppression. Fig. \ref{CMA_Even}(b) shows the |MWC| of the significant modes of the even spiral. The Mode e1 corresponds to the Mode t1 of the tube. It is evident that the spiral structure effectively suppresses Mode e1 over higher frequencies. However, it introduces Modes e2, e3, and e4, which limit the suppression bandwidth. To widen the suppression bandwidth, Modes e2 and e3 need to be removed or shifted to higher frequencies. The modal E-field distributions of Mode e2 and Mode e3 are shown in Fig. \ref{CMA_Even}(c). For Mode e2, the E-field is much stronger in the middle part and the two regions near the ends of the slot. For Mode e3, the E-field is also stronger in the middle part. These two modes can be eliminated by short-circuiting the slots in the areas with strong E-fields, which are marked with dash lines in Fig. \ref{CMA_Even}(c).

\begin{figure}[!t]
\centering
\includegraphics[trim=0 10 0 0,scale=1]{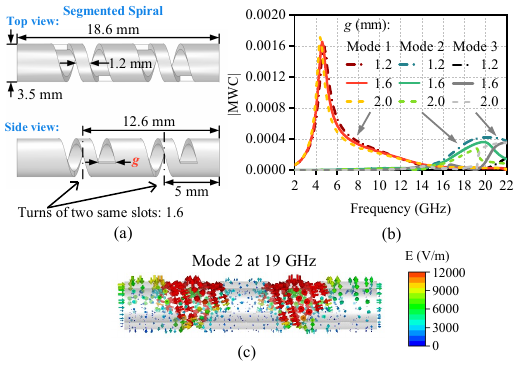}
\caption{(a) Geometry of the segmented spiral, (b) |MWC| of the segmented spiral with different width of slots (\textit{g}), and (c) modal E-field distribution of Mode 2.}
\label{CMA_Uneven}
\end{figure}

Short-circuiting the even spiral leads to a segmented spiral shown in Fig. \ref{CMA_Uneven}(a). As given in Fig. \ref{CMA_Uneven}(b), Modes e2 and e3 in Fig. \ref{CMA_Even}(b) are eliminated successfully, however, another higher-order mode named Mode 2 appears. Fig. \ref{CMA_Uneven}(c) presents its modal E-field distribution at 19 GHz. The areas with the strongest E-field are located around the two slots. Instead of completely eliminating this mode, a more feasible approach is to suppress the Mode 2 by modifying the slot widths of the two segmented spirals. As the slot width (\textit{g}) increases from 1.2 mm to 2.0 mm, the |MWC| of Mode 2 shown in Fig. \ref{CMA_Uneven}(b) gradually decreases and the |MWC| of Mode 1 at higher frequencies  is also slightly reduced, thereby widening the suppression bandwidth.

\begin{figure}[!t]
\centering
\includegraphics[trim=0 10 0 0,scale=1]{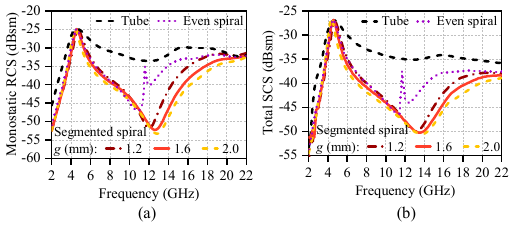}
\caption{(a) Monostatic RCSs and (b) total SCSs of the tube, the even spiral, and the segmented spiral.}
\label{RCS_SCS}
\end{figure}

\begin{figure}[!t]
\centering
\includegraphics[trim=0 10 0 0,width=8.8cm]{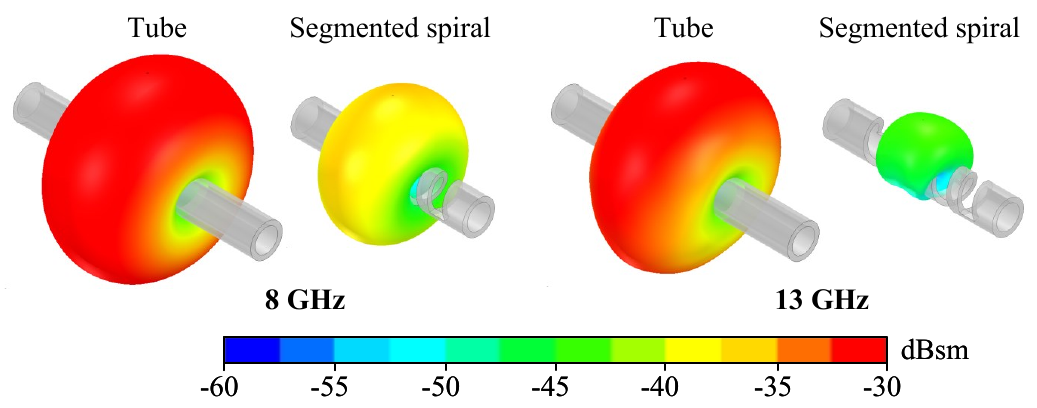}
\caption{Bistatic RCSs of the tube and the segmented spiral at two sample frequencies.}
\label{bi_RCS}
\end{figure}

To demonstrate the advantage of the segmented spiral in scattering suppression, its monostatic radar cross-sections (RCSs) and total scattering cross-sections (SCSs) are compared to those of the reference tube and the even spiral in Fig. \ref{RCS_SCS}. The monostatic RCS quantifies the backscattering intensity of the target in the direction coincident with the incident wave, whereas the total SCS represents the integration of the scattered power over all directions, providing an overall evaluation of the scattering intensity. As observed from Fig. \ref{RCS_SCS}, the even spiral reduces the RCS and SCS but is effective only within a limited frequency range. In contrast, the segmented spiral effectively reduces both the RCS and SCS over a wide bandwidth of 4.7–21.5 GHz (128.2\%) when \textit{g} = 1.6 mm, achieving maximum reductions of 20 dB in RCS and 15 dB in SCS.

Fig. \ref{bi_RCS} depicts the bistatic RCSs of the tube and the segmented spiral at 8 GHz and 13 GHz, intuitively illustrating their scattering at various observation angles. The segmented spiral exhibits significant scattering reduction over all directions compared with the tube. It can be concluded that employing the segmented spiral as the LB antenna radiator enables restoration of the distorted MB/HB radiation patterns across all directions.

\begin{figure}[!t]
\centering
\includegraphics[trim=0 10 0 0,scale=1]{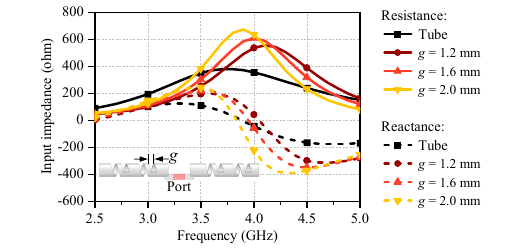}
\caption{Input impedance of the LB dipole using the segmented spiral or the tube.}
\label{input_impedance}
\end{figure}

Although a larger \textit{g} results in a wider suppression bandwidth, as shown in Fig. \ref{CMA_Uneven}(b) and Fig. \ref{RCS_SCS}, it is necessary to consider the scattering suppression and matching characteristic of the segmented spiral at the same time. Input impedance of the LB dipole using the segmented spiral or the tube is shown in Fig. \ref{input_impedance}. Increasing \textit{g} makes the input resistance and reactance curves steeper, which leads to increased difficulty in LB impedance matching. With an overall consideration of the scattering suppression and matching, \textit{g} = 1.6 mm is selected for the final design.

\subsection{LB Balun Design for Impedance Matching}

\begin{figure}[!t]
\centering
\includegraphics[trim=0 10 0 0,width=8.8cm]{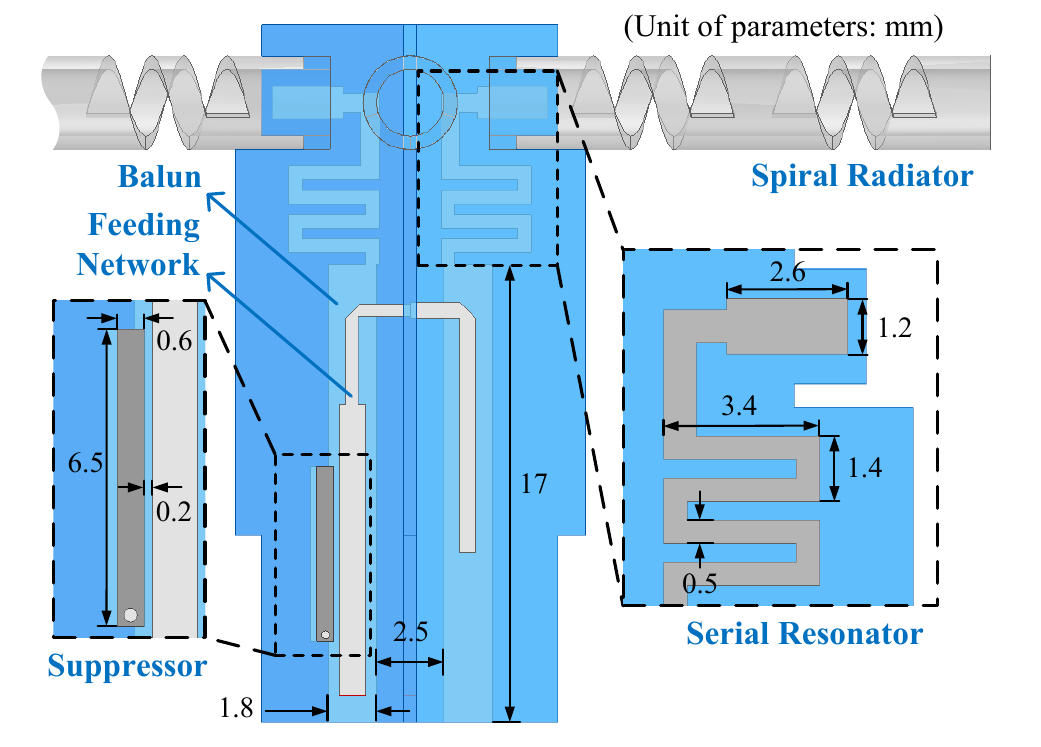}
\caption{Geometry of the LB segmented spiral antenna with integrated serial resonators and suppressors in the balun.}
\label{Balun}
\end{figure}

As shown in Fig. \ref{input_impedance}, the segmented spiral exhibits significantly larger variation in input impedance compared to the reference tube, posing a greater challenge for impedance matching. Simultaneously achieving wideband scattering suppression and wideband impedance matching is highly challenging, as there is a trade-off between them. To address this problem, a pair of serial resonators, as shown in Fig. \ref{Balun}, are added between the radiators and the balun to restore the impedance matching. The resonators, balun and feeding network are printed on two sides of a substrate with a relative permittivity of 2.2 and a thickness of 0.508 mm. A metal pad is inserted into the spiral radiator at the feeding position, forming a capacitor with the metal structure of the radiator. A meander line is added between the capacitor and the balun, serving as an inductor. This resonator, oriented orthogonally to the radiation polarization, features a coupled, non-contact loading to the spiral radiator, thereby avoiding the degradation of scattering suppression performance. In addition, a suppressor consisting of strip, vias and ground is positioned adjacent to the balun, which is used to improve the isolation between the LB and MB antennas in MB.

\begin{figure}[!t]
\centering
\includegraphics[trim=0 10 0 0,scale=1]{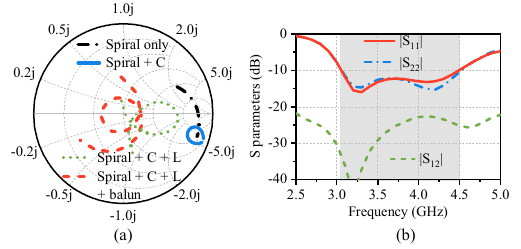}
\caption{(a) Input impedances of the LB segmented spiral dipole with different loading conditions. (b) Reflection coefficients (|S$_{11}$| and |S$_{22}$|), and transmission coefficient |S$_{12}$| of the LB segmented spiral antenna.}
\label{Smith}
\end{figure}

Fig. \ref{Smith}(a) plots the input impedances of the segmented spiral dipole with different loading conditions, illustrating how the serial resonators help restore impedance matching. The frequency range shown in the Smith chart is 3.0-4.9 GHz. The input impedance curve of “Spiral only” is far away from the matching point of the Smith chart, and the corresponding radius of the arc is very large. Therefore, it is difficult to achieve wide-band impedance matching using the conventional balun. To reduce the radius of the “Spiral only” arc, the capacitor is added, and the arc is rotated counterclockwise along the equal resistance circle in the Smith chart, obtaining a “Spiral + C” arc with a much smaller radius. The smaller the area of the metal pad constituting the capacitor, the larger the rotation angle and the smaller the radius of the arc. However, it is further away from the matching point. The addition of the inductor makes the arc rotate clockwise. The arc of “Spiral + C + L” is closer to the matching point with the increase in length of the meander line, indicating the potential for matching. Finally, a conventional balun with feeding network is introduced so that the “Spiral + C + L + balun” arc can wrap around the matching point with a small radius. The entire tuning process only adds the adjustment of two parameters, the capacitance and the inductance, to rotate the arcs on the Smith chart into a characteristic fish-like pattern indicating multiple resonances, as shown in Fig. \ref{Smith}(a). A detailed procedure for achieving impedance matching using the additional LC structure can be found in our previous work \cite{Full}.

The S parameters of the matched LB spiral antenna are presented in Fig. \ref{Smith}(b), and ports 1 and 2 represent the two ports of the LB cross-dipole antenna shown in Fig. \ref{Balun}. The reflection coefficients (|S$_{11}$| and |S$_{22}$|) are less than -10 dB in 3.05-4.51 GHz, and the polarization isolation exceeds 20 dB within this range. The wideband matching performance validates the efficacy of the serial-resonator-integrated balun design.

\section{MB and HB Antenna Design}

\begin{figure}[!t]
\centering
\includegraphics[trim=0 10 0 0,width=8.8cm]{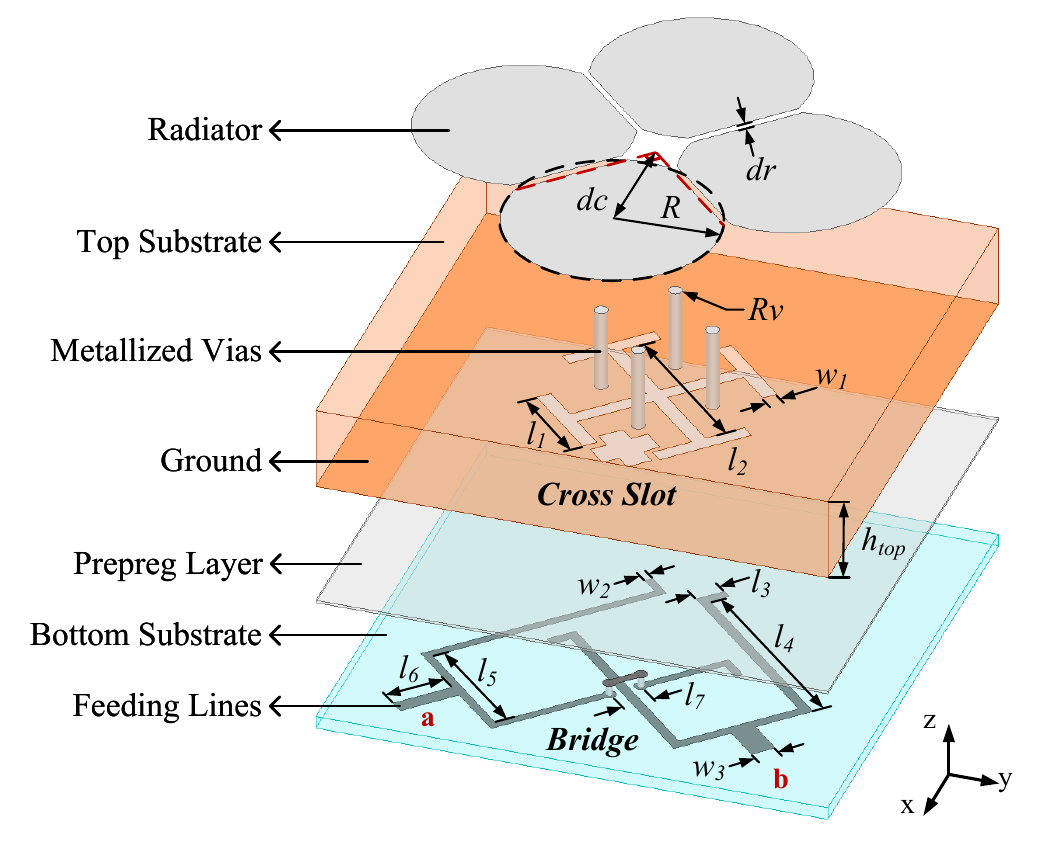}
\caption{Geometry of the MB and HB planar ME dipole antenna.}
\label{MB_HB}
\end{figure}

The anticipated band for 5G-Advanced and 6G is 6.425–15.35 GHz. The required bandwidth is very wide and the frequency is significantly higher than that of 3G, 4G and 5G (sub-6 GHz). If conventional cross-dipole is still used, the designed antenna size will be extremely small, posing significant challenges for fabrication and assembly. To achieve the required wide operational bandwidth, the planar magnetoelectric (ME) dipole structure \cite{ME1}, \cite{ME2} is a strong candidate.

\begin{table}
\centering
\renewcommand{\arraystretch}{1}
\caption{Geometrical Parameters of the MB and HB Antennas}
\label{table:params}
\vspace{-3pt}
\begin{tabular}{
>{\centering\arraybackslash}p{0.7cm} 
>{\centering\arraybackslash}p{0.7cm} 
>{\centering\arraybackslash}p{0.7cm} 
>{\centering\arraybackslash}p{0.7cm} 
>{\centering\arraybackslash}p{0.7cm} 
>{\centering\arraybackslash}p{0.7cm} 
>{\centering\arraybackslash}p{0.7cm}
}
\toprule[1pt]
\specialrule{0em}{0.4pt}{0.4pt}
\toprule[1pt]
\textbf{}   & \textit{\textbf{R}}   & \textit{\textbf{dc}} & \textit{\textbf{dr}}  & \textit{\textbf{Rv}}  & \textit{\textbf{h\textsubscript{top}}} & \textit{\textbf{l\textsubscript{1}}}  \\ 
\midrule[0.8pt]
\textbf{MB} & 4.2                   & 5                    & 0.35                  & 0.25                  & 3.43                   & 3.8                   \\
\textbf{HB} & 2.65                  & 3                    & 0.37                  & 0.2                   & 2.54                   & 2.4                   \\
\midrule[1.2pt]
\textbf{}   & \textit{\textbf{l\textsubscript{2}}}  & \textit{\textbf{l\textsubscript{3}}} & \textit{\textbf{l\textsubscript{4a}}} & \textit{\textbf{l\textsubscript{4b}}} & \textit{\textbf{l\textsubscript{5a}}}  & \textit{\textbf{l\textsubscript{6a}}} \\
\midrule[0.8pt]
\textbf{MB} & 6.9                   & 1.2                  & 9.7                   & 8.5                   & 5                      & 2.5                   \\
\textbf{HB} & 4.4                   & 0.75                 & 6                     & 6                     & 3.2                    & 1.6                   \\
\midrule[1.2pt]
\textbf{}   & \textit{\textbf{l\textsubscript{6b}}} & \textit{\textbf{l\textsubscript{7}}} & \textit{\textbf{w\textsubscript{1}}}  & \textit{\textbf{w\textsubscript{2}}}  & \textit{\textbf{w\textsubscript{3a}}}  & \textit{\textbf{w\textsubscript{3b}}} \\
\midrule[0.8pt]
\textbf{MB} & 1.6                   & 1.3                  & 0.6                   & 0.35                  & 0.6                    & 1                     \\
\textbf{HB} & 1.6                   & 1.2                  & 0.4                   & 0.3                   & 0.6                    & 0.6                   \\
\bottomrule [1pt]\specialrule{0em}{0.4pt}{0.4pt}\bottomrule [1pt]
\end{tabular}
\end{table}

The designed planar ME dipole antenna structure, serving as the MB/HB elements in the tri-band array, is shown in Fig. \ref{MB_HB}. The four-leaf-clover-shaped radiator is connected to the ground through four vias, while the ±45° oriented fork-shaped microstrip lines excite the antenna through the Jerusalem-cross slots in the ground. To mitigate and avoid the coupling between the two polarizations of the antenna, the slot reserved for the bridge on the ground is designed in a cross shape, and the microstrip lines in the two directions have different lengths. The top and bottom substrate layers use materials with thickness of \textit{h\textsubscript{top}} (relative permittivity of 2.2) and 0.406 mm (relative permittivity of 3.55), respectively. A prepreg layer with a thickness of 0.102 mm is inserted between them during the lamination process. The other parameter values of the MB and HB antennas are provided in Table I. Note that they share a similar structure but have different dimension values.

\begin{figure}[!t]
\centering
\includegraphics[trim=0 10 0 0,width=8.8cm]{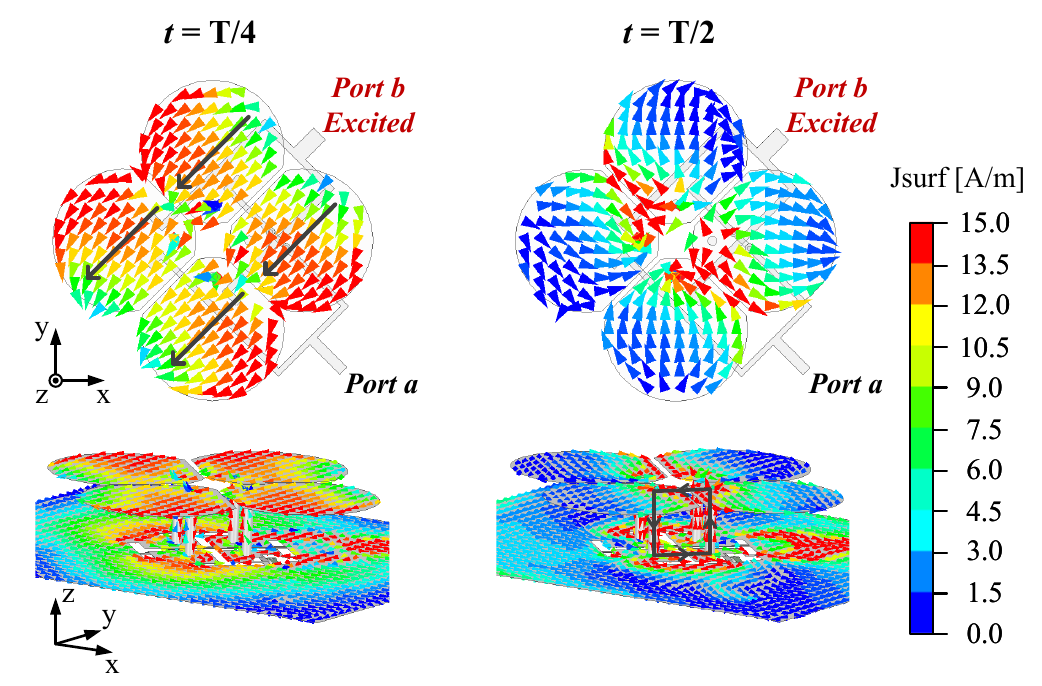}
\caption{Current distributions at 8.2 GHz on the MB ME dipole antenna at \textit{t} = T/4 and \textit{t} = T/2.}
\label{Current_ME}
\end{figure}

The four-leaf-clover-shaped radiator serves as the electric dipole of the designed ME dipole antenna and, together with four metallized vias and the metallic ground, forms the magnetic dipole of the ME dipole antenna. Current distributions at 8.2 GHz on the MB ME dipole antenna are shown in Fig. \ref{Current_ME}. At \textit{t} = T/4, in-phase currents are excited on the four-leaf-clover-shaped radiator, which serves as an electric dipole. At \textit{t} = T/2, strong currents are excited on the four-leaf-clover-shaped radiator, the metallized vias, and the metallic ground, forming a current loop that acts as a magnetic dipole. At \textit{t} = 3T/4 and \textit{t} = T, the current exhibits the same magnitude distributions as at \textit{t} = T/4 and \textit{t} = T/2, respectively, while the current directions are reversed. These current distributions agree with the characteristics of the ME dipole antenna \cite{ME2}, and the HB ME dipole will exhibit the same effect.

In a tri-band antenna array, employing planar ME dipoles as the MB and HB antenna elements can avoid common-mode resonance \cite{S12} within the LB and MB, due to their significantly lower profile compared to traditional 3D dipole antennas. Moreover, the MB and HB radiators are positioned at almost identical heights, effectively mitigating the cross-band scattering \cite{coplanar1}, \cite{coplanar2}, \cite{coplanar3} from the MB radiators in the HB.

\begin{figure}[!t]
\centering
\includegraphics[trim=0 10 0 0,scale=1]{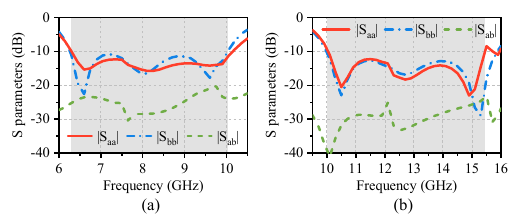}
\caption{Reflection coefficients (|S\textsubscript{aa}| and |S\textsubscript{bb}|), and transmission coefficient |S\textsubscript{ab}| of the (a) MB and (b) HB antenna.}
\label{MB_HB_Unit_S}
\end{figure}

\begin{figure}[!t]
\centering
\includegraphics[trim=0 10 0 0,scale=1]{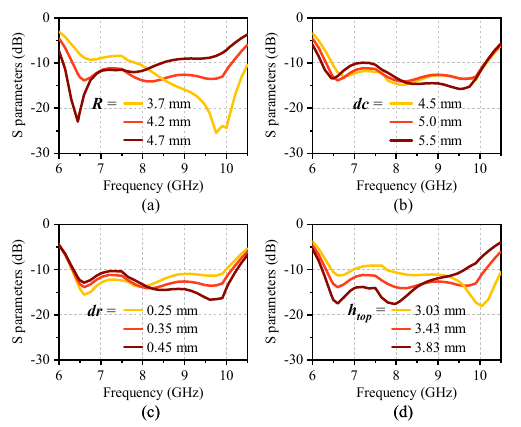}
\caption{Parametric analysis of the four-leaf-clover-shaped radiator on the reflection coefficients (|S\textsubscript{bb}|): (a) radius of the arc \textit{R}, (b) position of the arc centre \textit{dc}, (c) spacing between the elements \textit{dr}, and (d) height of the radiator \textit{h\textsubscript{top}}.}
\label{Sweep_parameters_MB}
\end{figure}

According to the simulation results shown in Fig. \ref{MB_HB_Unit_S}, the proposed MB and HB planar ME dipoles work in the bandwidths of 6.3-10.0 GHz and 10.0-15.45 GHz, respectively, covering the 5G-Advanced and 6G range of 6.425–15.35 GHz. Their polarization isolations exceed 20 dB across the operating bands. The wideband impedance matching is attributed to the designed four-leaf-clover-shaped radiator. Fig. \ref{Sweep_parameters_MB} presents the effects of its structural parameters on the reflection coefficient of the ME dipole, using the MB antenna as an example. The radius of the arc (\textit{R}) determines the size of the radiator and thus defines the overall operating band of the antenna. The position of the arc centre (\textit{dc}) affects the lowest operating frequency, whereas the spacing between the four elements (\textit{dr}) influences the highest operating frequency. Moreover, the height of the radiator (\textit{h\textsubscript{top}}) impacts the matching bandwidth and the matching level of the antenna.

\begin{figure}[!t]
\centering
\includegraphics[trim=0 10 0 0,scale=1]{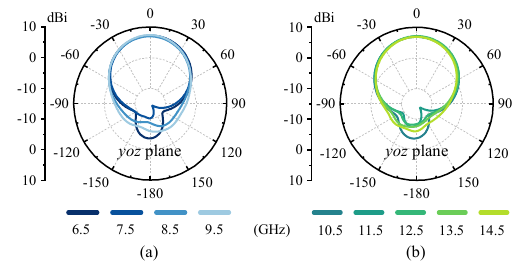}
\caption{Radiation patterns of the (a) MB and (b) HB antenna.}
\label{MB_HB_Unit_Patterns}
\end{figure}

The radiation patterns of the MB and HB antennas in the \textit{yoz} plane are shown in Fig. \ref{MB_HB_Unit_Patterns}, which are stable and exhibit good consistency over their respective bands. The radiation patterns of the MB and HB ME dipoles show half-power beamwidths of $74.5^\circ \pm 2.5^\circ$ and $78.5^\circ \pm 3.5^\circ$, respectively, which are comparable to and slightly wider than those of conventional dipoles, whose half-power beamwidths are generally around $60^\circ$–$70^\circ$ \cite{HPBW1}, \cite{HPBW2}. The increased beamwidth is attributed to the lower profile of the ME dipoles compared to standard dipoles, but it remains pretty stable across the wide operating band.

\section{Holistic Suppression of Scattering and Coupling in The Tri-Band Array}

\begin{figure}[!t]
\centering
\includegraphics[trim=0 10 0 0,width=8.8cm]{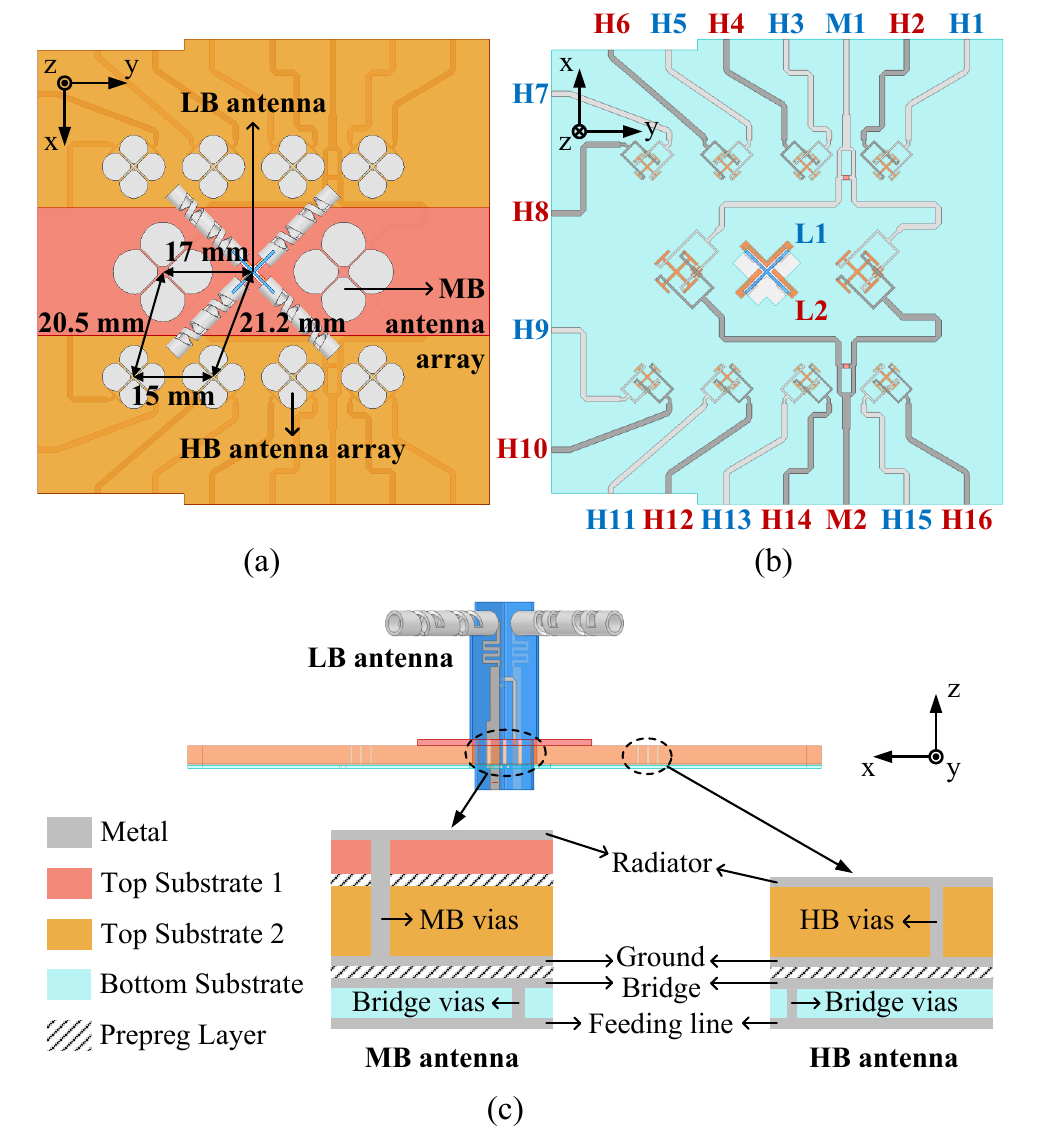}
\caption{(a) Top view, (b) bottom view, and (c) side view of the tri-band antenna array.}
\label{Tri_Band_Array_All_A}
\end{figure}

As illustrated in Fig. \ref{Tri_Band_Array_All_A}, the proposed LB, MB, and HB antennas are compactly arranged in an interleaved tri-band array to validate the scattering and coupling suppression performance of the proposed methods and structures. The bottom-substrate thickness of the MB and HB antennas is identical, but the top-substrate thickness differs to enable wider operating bandwidths. Therefore, the MB and HB antenna arrays are designed as a planar structure composed of three laminated dielectric layers, with a slot etched in the central region to accommodate the LB antenna. The minimum center-to-center spacings between the LB and MB elements, LB and HB elements, and MB and HB elements are 17 mm (0.22 $\lambda_{\text{LB}}$), 21.2 mm (0.27 $\lambda_{\text{LB}}$), and 20.5 mm (0.55 $\lambda_{\text{MB}}$), respectively.

\begin{figure}[!t]
\centering
\includegraphics[trim=0 10 0 0,scale=1]{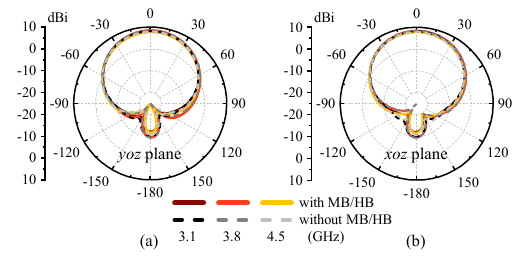}
\caption{Simulated radiation patterns of the LB antenna in (a) \textit{yoz} plane and (b) \textit{xoz} plane, when port L1 is excited.}
\label{Array_LB_Patterns_Ref}
\end{figure}

Fig. \ref{Array_LB_Patterns_Ref} compares the simulated radiation patterns of the LB antenna in the presence and absence of the MB and HB antennas. Owing to the proposed planar ME dipole structure, the common-mode resonance that could be induced by the MB and HB antennas is effectively avoided within the LB. Consequently, the LB antenna, surrounded by the MB and HB antennas, maintains undistorted radiation patterns over the entire band, in good agreement with those of the standalone LB antenna.

\begin{figure}[!t]
\centering
\includegraphics[trim=0 10 0 0,scale=1]{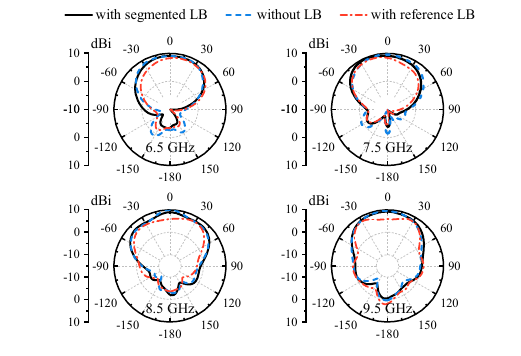}
\caption{Simulated radiation patterns of the MB antennas with and without the presence of different LB antennas, when port M1 is excited.}
\label{Array_MB_Patterns}
\end{figure}

\begin{figure}[!t]
\centering
\includegraphics[trim=0 10 0 0,scale=1]{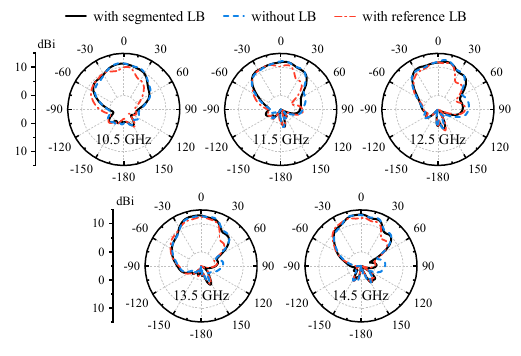}
\caption{Simulated radiation patterns of the HB antennas with and without the presence of different LB antennas, when ports H1-4 are excited.}
\label{Array_HB_Patterns}
\end{figure}

The MB and HB radiation patterns in the \textit{xoz} plane under different cases are compared in Fig. \ref{Array_MB_Patterns} and Fig. \ref{Array_HB_Patterns}, respectively. In the reference case, the conventional tube-based LB antenna significantly distorts the MB/HB radiation patterns and degrades the realized gain. In contrast, employing the modified LB antenna with MB/HB-transparent segmented spirals results in minimal impact on the MB/HB radiation patterns and realized gain, which remain closely aligned with those of the MB/HB antennas operating independently, without the LB antenna.

\begin{figure}[!t]
\centering
\includegraphics[trim=0 10 0 0,scale=1]{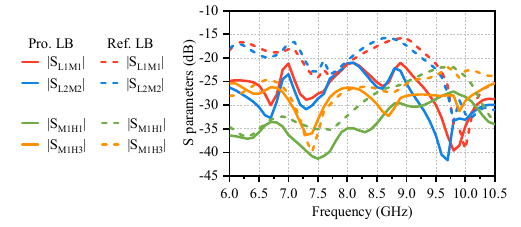}
\caption{Transmission coefficients in the MB between the LB, MB, and HB antennas.}
\label{Array_MB_iso}
\end{figure}

\begin{figure}[!t]
\centering
\includegraphics[trim=0 10 0 0,scale=1]{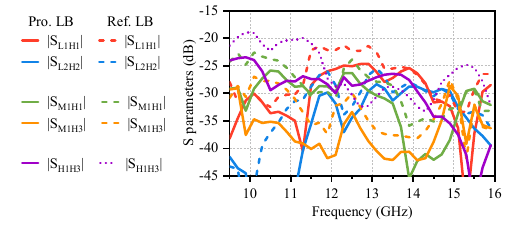}
\caption{Transmission coefficients in the HB between the LB, MB, and HB antennas.}
\label{Array_HB_iso}
\end{figure}

In addition to effectively suppressing cross-band scattering and restoring radiation patterns, the proposed suppression structures are also capable of mitigating coupling, thereby enhancing the isolation in the tri-band antenna array. Fig. \ref{Array_MB_iso} and Fig. \ref{Array_HB_iso} show representative isolations between the LB, MB, and HB antennas in the tri-band array employing either the proposed LB antenna or the reference tube-radiator LB antenna. The proposed LB antenna achieves higher isolations in both the MB and HB compared to the reference case. The MB/HB-transparent segmented-spiral-based LB radiator induces and re-radiates less MB/HB energy than the reference tube radiator, resulting in reduced induced currents flow to the LB ports and weaker coupling to the other MB/HB antennas. In addition, the suppressor in the LB balun further blocks the MB induced currents from entering the LB ports. As a result, the cross-band coupling in the MB is reduced to below -20 dB, while both the cross-band and in-band coupling in the HB are suppressed to around -25 dB.

\section{Results and Discussions}
\subsection{Experiment Results}

\begin{figure}[!t]
\centering
\includegraphics[trim=0 15 0 0,width=8.8cm]{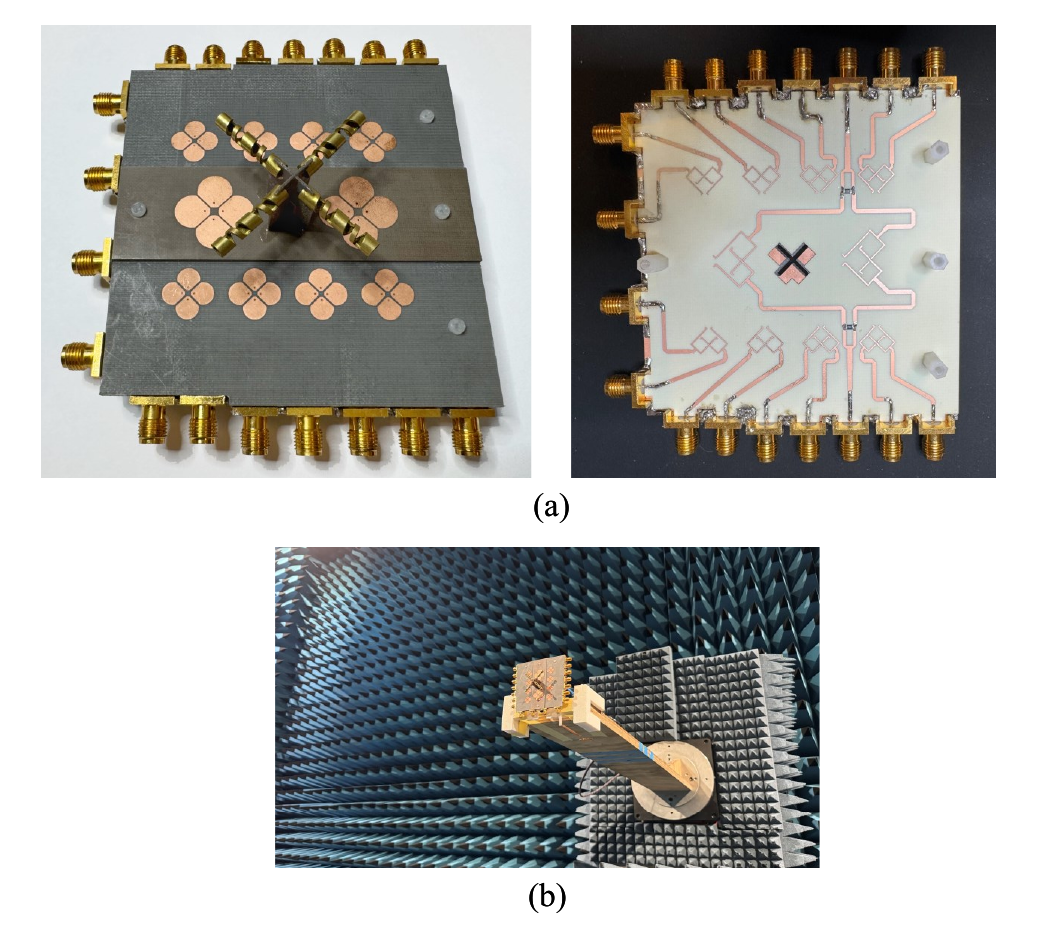}
\caption{(a) Prototype of the tri-band antenna array and (b) far-field test environment.}
\label{Tri_Band_Array_All_B}
\end{figure}

\begin{figure}[!t]
\centering
\includegraphics[trim=0 10 0 0,scale=1]{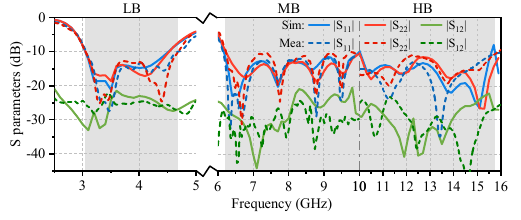}
\caption{Simulated and measured S parameters of the LB, MB, and HB antennas in the tri-band array.}
\label{Array_LB_MB_HB_S}
\end{figure}

The prototype of the proposed tri-band antenna array and the far-field test environment are shown in Fig. \ref{Tri_Band_Array_All_B}(a) and (b), respectively. Fig. \ref{Array_LB_MB_HB_S} shows the simulated and measured S parameters of the antennas. The measured reflection coefficients (|S$_{11}$| and |S$_{22}$|) of the LB, MB, and HB antennas are less than -10 dB in 3.05-4.68 GHz (42.2\%), 6.2-10.0 GHz (46.9\%), and 10.0-15.6 GHz (43.8\%), respectively, which agree well with the simulated ones. Meanwhile, each of the three antenna types maintains over 20 dB polarization isolation across its respective operating band.

\begin{figure}[!t]
\centering
\includegraphics[trim=0 10 0 0,scale=1]{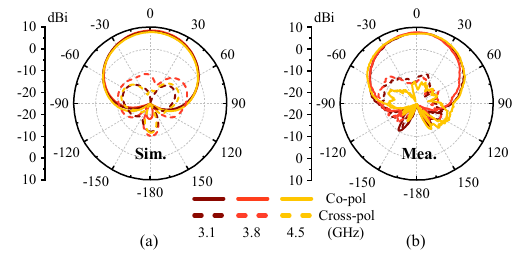}
\caption{(a) Simulated and (b) measured radiation patterns of the LB antenna in \textit{yoz} plane, when port L1 is excited.}
\label{Array_LB_Patterns}
\end{figure}

\begin{figure}[!t]
\centering
\includegraphics[trim=0 10 0 0,scale=1]{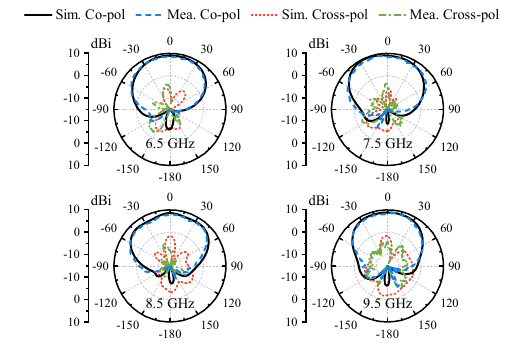}
\caption{Simulated and measured radiation patterns of the MB antennas in the array.}
\label{Mea_Array_MB_Patterns}
\end{figure}

The simulated and measured radiation patterns of the LB antenna are compared in Fig. \ref{Array_LB_Patterns}. The measured radiation patterns show good agreement with the simulated results, both exhibiting no significant distortion. These results confirm the effectiveness of the proposed planar ME dipole structure for the MB and HB antennas in avoiding the common-mode resonance commonly occurring in the LB.

\begin{figure}[!t]
\centering
\includegraphics[trim=0 10 0 0,scale=1]{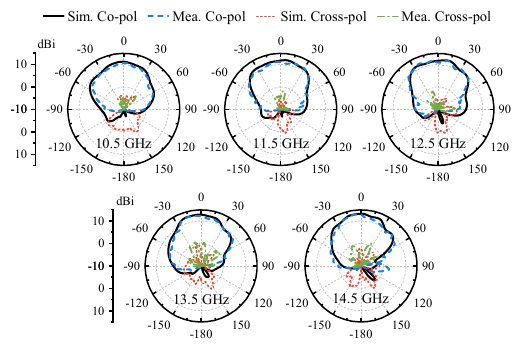}
\caption{Simulated and measured radiation patterns of the HB antennas in the array.}
\label{Mea_Array_HB_Patterns}
\end{figure}

The measured radiation patterns of the MB and HB antennas in the \textit{xoz} plane are shown in Figs. \ref{Mea_Array_MB_Patterns} and \ref{Mea_Array_HB_Patterns}, respectively. The measured radiation patterns agree well with the simulated results. These undistorted MB/HB patterns validate the efficacy of the designed MB/HB-transparent segmented spiral in suppressing cross-band scattering across the ultra-wide frequency bands.

\begin{figure}[!t]
\centering
\includegraphics[trim=0 10 0 0,scale=1]{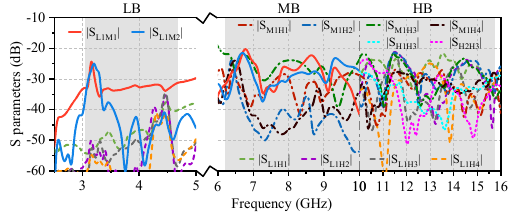}
\caption{Measured transmission coefficients between the LB, MB, and HB antennas in the tri-band antenna array.}
\label{Array_LB_MB_HB_iso}
\end{figure}

The measured isolation between any two ports in the proposed tri-band antenna array exceeds 20 dB across the three target bands, and representative results are shown in Fig. \ref{Array_LB_MB_HB_iso}. Therefore, the proposed suppression techniques not only achieve undistorted radiation patterns but also ensure good cross-band and in-band isolation over wide bands, holistically suppressing both scattering and coupling in the tri-band antenna array.

\subsection{Beam Scanning Performance}

\begin{figure}[!t]
\centering
\includegraphics[trim=0 10 0 0,width=8.8cm]{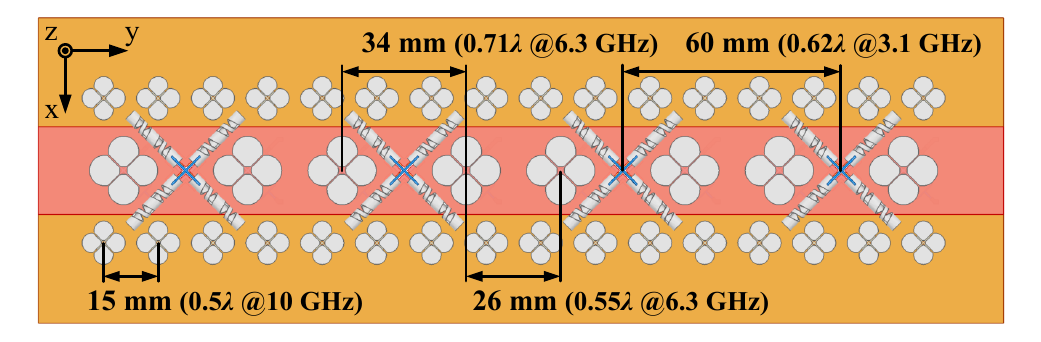}
\caption{Tri-band antenna array for beam scanning.}
\label{Four_array}
\end{figure}

\begin{figure}[!t]
\centering
\includegraphics[trim=0 5 0 0,scale=1]{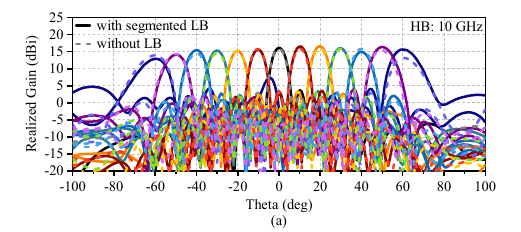}
\includegraphics[trim=0 10 0 0,scale=1]{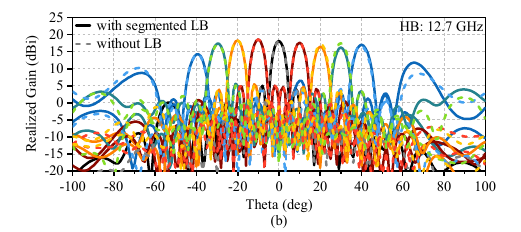}
\caption{Beam scanning performance of the HB antenna array at (a) 10 GHz, and (b) 12.7 GHz.}
\label{Beam_scanning_HB}
\end{figure}

\begin{figure}[!t]
\centering
\includegraphics[trim=0 10 0 0,scale=1]{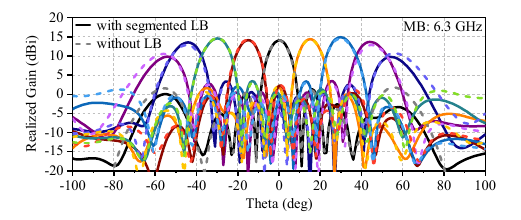}
\caption{Beam scanning performance of the MB antenna array at 6.3 GHz.}
\label{Beam_scanning_MB}
\end{figure}

\begin{figure}[!t]
\centering
\includegraphics[trim=0 10 0 5,scale=1]{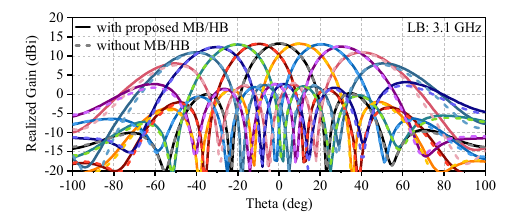}
\caption{Beam scanning performance of the LB antenna array at 3.1 GHz.}
\label{Beam_scanning_LB}
\end{figure}

Although only one section of the tri-band array was fabricated, the design can be readily scaled up to a larger array for practical applications, as illustrated in Fig. \ref{Four_array}. The simulated beam scanning performance in the HB, MB, and LB bands is shown in Figs. \ref{Beam_scanning_HB}–\ref{Beam_scanning_LB}, respectively. The element spacings of the LB, MB, and HB antennas are labelled in the figure. With the given spacing, all the isolations remain over 20 dB. The scanning method employs the simplest excitation scheme, in which all antenna elements are excited with equal amplitudes, and adjacent elements have identical phase differences.

As shown in Fig. \ref{Beam_scanning_HB}(a), when the 45°-polarized HB elements in the bottom row are excited, the HB beam can be scanned up to $\pm60^{\circ}$ at the lowest HB frequency of 10 GHz with a realized gain reduction of approximately 3 dB. However, at the higher frequency of 12.7 GHz, the maximum scanning angle decreases to $\pm40^{\circ}$, as shown in Fig. \ref{Beam_scanning_HB}(b), because the scanning range is constrained by the relatively larger element spacing. The scanning performances of the 45°-polarized MB and LB arrays are presented in Fig. \ref{Beam_scanning_MB} and Fig. \ref{Beam_scanning_LB}, respectively. Their beams can be scanned to $\pm45^{\circ}$ at 6.3 GHz for the MB array and $\pm40^{\circ}$ at 3.1 GHz for the LB array. The HB, MB, and LB scanning beams exhibit no significant distortion and agree well with those obtained without the presence of the LB or MB/HB antennas. This is attributed to the designed LB antenna with strong scattering-suppression capability in the MB and HB, as well as to the planar MB/HB antennas that avoid common-mode resonances in the LB.

\subsection{Comparison and Discussion}

To demonstrate the advantages of techniques developed in this work, their performance is compared with that of other techniques for scattering and coupling suppression in tri-band shared-aperture array, as listed in Table II. This work achieves tri-functional suppression of scattering and coupling without increasing the complexity of the array configuration, featuring the widest reported bandwidths across all three bands: LB, MB, and HB.

\begin{table*}
 \centering
 \label{table1}
 \caption{Comparison of Techniques for Scattering and Coupling Suppression in Tri-Band Antenna Arrays}
\vspace{-5pt}

 \centering
 \begin{tabular}{cccccccc}
\toprule[1.2pt]
\specialrule{0em}{0.5pt}{0.5pt}
\toprule[1.2pt]
\multirow{2}{*}{\textbf{Ref.}} & 
\multirow{2}{*}{\textbf{\begin{tabular}[c]{@{}c@{}}Proposed\\structures\end{tabular}}} & 
\multirow{2}{*}{\textbf{\begin{tabular}[c]{@{}c@{}}Suppression\\functions\end{tabular}}} & 
\multirow{2}{*}{\textbf{\begin{tabular}[c]{@{}c@{}}Operating band\\(GHz)\end{tabular}}} & 
\multirow{2}{*}{\textbf{\begin{tabular}[c]{@{}c@{}}Cross-band\\isolation (dB)\end{tabular}}} & 
\multirow{2}{*}{\textbf{\begin{tabular}[c]{@{}c@{}}In-band\\isolation (dB)\end{tabular}}} 
\\
                               &                                                                                        & \begin{tabular}[c]{@{}c@{}}\end{tabular} &                                                                                          &                                                                                                 &                                                                                              \\ \midrule [1.2pt]

\cite{T5}                       & LB: Dual-band FSS radiator                                                                  & Cross-band
scattering
                                                                                                                                                                                      & LB: 1.85-2.15 (15\%)                                                                   & N/A     & N/A  \\
                               & MB/HB: DRA        &                                                                    & MB: 3.4-3.6 (5.7\%)  &                                                                                                 &  \\  & & & HB: 5.4-5.6 (3.6\%) & &
\\  \midrule [1pt]

\cite{T1}                        & LB: Choke + Slot                                                                  & Cross-band
scattering
                                                                                                                                                                                      & LB: 0.79-0.96 (19.4\%)                                                                   & N/A     & N/A  \\
                               & MB: Slot + Metal strip &                                                                    & MB: 1.71-2.17 (23.7\%)                                                                   &                                                                                                 &                \\ & & & HB: 3.4-3.6 (5.7\%) & &
\\  \midrule [1pt]

\cite{T3}                        & LB: 45\degree \ rotation + slot                                                                  & Cross-band
scattering
                                                                                                                                                                                      & LB: 0.76-0.88 (14.6\%)                                                                   & N/A     & N/A  \\
                               & MB: FSS radiator &                                                                    & MB: 1.9–2.7 (34.8\%)   \\& & & HB: 3.3–3.9 (16.7\%) & &                                                                  &                                                                                                 &
\\  \midrule [1pt]

\cite{T2}                        & Stacked arrangement with FSS                                                                  & Cross-band
scattering
                                                                                                                                                                                      & LB: 0.69-0.96 (32.7\%)                                                                   & \textgreater \ 20 (LB)     & \textgreater \ 20 (HB)  \\
                               & MB: FSS radiator + Helical cable & Cross-band coupling                                                                   & MB: 1.8–2.7 (40.0\%)                                                                  &  \textgreater \ 19 (MB)                                                                                               &                         \\
     & HB: Fence + Helical cable        & In-band coupling                                                                   & HB: 3.3–3.8 (14.1\%)                                                                  &  \textgreater \ 19 (HB)                                                                                               &                                                    
\\  \midrule [1pt]

\cite{T7}                       & LB/MB: A dual-band antenna                                                                   & Cross-band
scattering
                                                                                                                                                                                      & LB/MB: 1.427-1.518 (6.2\%)                                                                   & \textgreater \ 20 (LB/MB)     & N/A  \\
                               & with split rings              & Cross-band coupling                                                                   & \;\quad \qquad 1.92-2.18 (12.7\%)     &  \textgreater \ 20 (HB)                                                                                               &                         \\
     & HB: Stacked arrangement         &                                                                    & HB: 3.3–3.8 (14.1\%)                                                                  &    &                                                    
\\  \midrule [1pt]

This                           & LB: Segmented spiral                                                                           & Cross-band scattering & LB: 3.05-4.68 (\textbf{42.2\%})  & \textgreater \ 20 (LB)                                                                                        & \textgreater \ 20 (HB)                                                                                      \\
work                           &+ Suppressor    &   Cross-band coupling                                                                                                                     & MB: 6.2-10.0 (\textbf{46.9\%})
                                                                     & \textgreater \ 20 (MB)                                                                                         &              \\   & MB/HB: Planar ME dipole & In-band coupling      &      HB: 10.0-15.6 (\textbf{43.8\%})     & \textgreater \ 20 (HB)                                                                                                                                                              \\
                               \bottomrule [1.2pt]\specialrule{0em}{0.5pt}{0.5pt}\bottomrule [1.2pt]
\vspace{-18pt}                      
\end{tabular}
\end{table*}

It is also worth mentioning that the developed MB and HB antennas necessarily adopt a relatively complex three-layer laminated dielectric structure to cover the ultra-wideband 5G-Advanced and anticipated 6G frequency range of 6.425–15.35 GHz (82\%). Fortunately, the fabrication process for three-layer PCBs is now highly mature, offering low manufacturing tolerances and good scalability for mass production. However, this comes at a higher cost compared to single- or double-layer PCBs. Additionally, in base station deployments, such multilayer substrates may reduce forward thermal dissipation compared to single-layer designs. As 6G systems continue to evolve toward higher integration and complexity, improved thermal and power management solutions will be essential. One promising direction is to enhance backward thermal dissipation through approaches such as using a metal baseplate with a fin-type heat sink, or incorporating additional air- or liquid-cooling systems \cite{Thermal1}, \cite{Thermal2}, \cite{Thermal3}.

\section{Conclusion}

This work proposes a tri-band shared-aperture 5G/6G antenna array featuring wideband scattering and coupling suppression, which is achieved by a modified LB antenna with segmented spirals, serial resonators, and suppressors, along with MB and HB antennas based on planar ME dipole structures. The LB antenna (3.05–4.68 GHz, 42.2\%) covers the 5G band of 3.3–4.2 GHz, while the MB (6.2–10.0 GHz, 46.9\%) and HB (10.0–15.6 GHz, 43.8\%) antennas collectively span the expected 5G-Advanced and 6G spectrum of 6.425-15.35 GHz. Across all three bands, the antennas maintain undistorted radiation patterns, and the isolation between any two ports in the array exceeds 20 dB. The measured results show good agreement with the simulations, validating the effective suppression of scattering and coupling over the wide bands. Therefore, the proposed tri-band shared-aperture 5G/6G antenna array is a promising candidate for future 6G applications.

\bibliographystyle{IEEEtran}
\bibliography{Reference}

\begin{IEEEbiography}[{\includegraphics[width=1in,height=1.25in,clip,keepaspectratio]{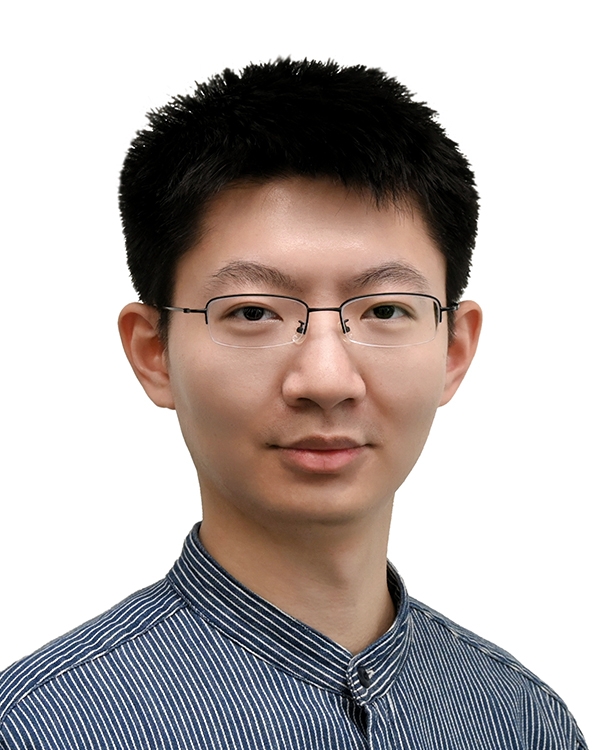}}]{Shang-Yi Sun}
(Graduate Student Member, IEEE) received the bachelor’s degree and the master’s degree from Xidian University, Xi’an, China, in 2016 and 2019, respectively. He is currently pursuing the Ph.D. degree in engineering with the Global Big Data Technologies Centre (GBDTC), University of Technology Sydney (UTS), NSW, Australia. His research interests include scattering and coupling suppression, multi-band shared-aperture antenna arrays, and base-station antennas. 

He was a spotlighted student and young professional at the IEEE Foundation Day in 2025. He was a recipient of the 2024 IEEE Antennas and Propagation Society C. J. Reddy Travel Grant for Graduate Students, and the Best Antenna Paper Award at the 2023 IEEE International Symposium on Antennas and Propagation. He was a Best Student Paper Award finalist at the 2024 International Symposium on Antennas and Propagation and the 2023 5th Australian Microwave Symposium. He is serving as a reviewer for multiple flagship journals, including \textit{IEEE Transactions on Antennas and Propagation}, \textit{IEEE Antennas and Wireless Propagation Letters}, and \textit{IEEE Open Journal of Antennas and Propagation}.
\end{IEEEbiography}

\vspace{-1.5\baselineskip}

\begin{IEEEbiography}[{\includegraphics[width=1in,height=1.25in,clip,keepaspectratio]{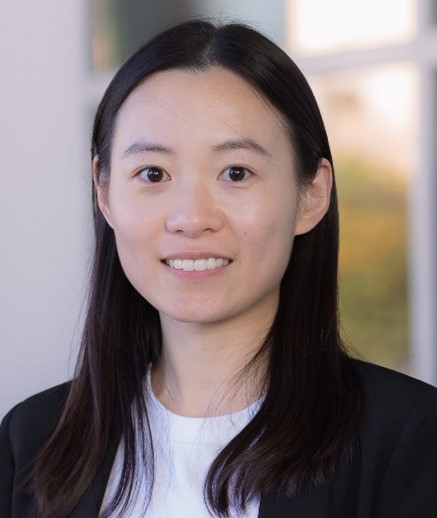}}]{Hai-Han Sun}
(Senior Member, IEEE) received her bachelor’s degree in electronic information engineering from Beijing University of Posts and Telecommunications, Beijing, China, in 2015, and the Ph.D. degree in engineering from the University of Technology Sydney, Australia, in 2019. From 2019 to 2023, she was a Research Fellow at Nanyang Technological University, Singapore. 

She is currently an Assistant Professor in the Department of Electrical and Computer Engineering at the University of Wisconsin-Madison. Her research interests include ground-penetrating radar, base station antenna, microwave sensing, and non-destructive testing. She was a recipient of several Young Professional and Student Paper awards at international conferences. She was awarded the Mojgan Daneshmand Grant for Women by the IEEE Antennas and Propagation Society (AP-S) in 2021. She was recognized as an Outstanding Associate Editor for both the \textit{IEEE Antennas and Wireless Propagation Letters} and the \textit{IEEE Open Journal of Antennas and Propagation} in 2025. She is currently serving as an Associate Editor for \textit{IEEE Transactions on Geoscience and Remote Sensing} and \textit{IEEE Antennas and Wireless Propagation Letters}, and is also an IEEE AP-S Young Professional Ambassador.
\end{IEEEbiography}

\vspace{-1.5\baselineskip}

\begin{IEEEbiography}[{\includegraphics[width=1in,height=1.25in,clip,keepaspectratio]{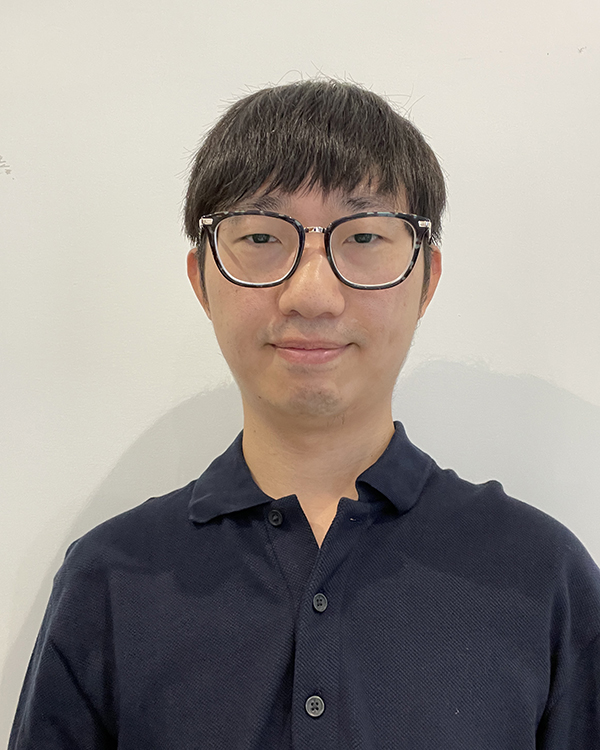}}]{Can Ding}
(Senior Member, IEEE) received the bachelor’s degree in integrated circuit and integrated system from Xidian University, Xi’an, China, in 2009, and the joint Ph.D. degree in electromagnetic fields and microwave technology from Xidian University and Macquarie University, Sydney, NSW, Australia, in 2016.

From 2015 to 2017, he was a Post-Doctoral Research Fellow with the University of Technology Sydney (UTS), Ultimo, NSW, Australia, where he is currently an Associate Professor with the Faculty of Engineering and IT (FEIT), and a Core Member of the Global Big Data Technologies Center (GBDTC). His accomplishments encompass several research and industry projects, patented innovations, and a portfolio of over 140 publications in top-tier journals and conferences. 

He has been an ARC DECRA Fellow from 2020 to 2024. He currently serves as an Associate Editor of the \textit{IEEE Transactions on Antennas and Propagation} (TAP) and the \textit{IEEE Antennas and Wireless Propagation Letters} (AWPL). He was recognized as a Top Reviewer for the IEEE TAP for four consecutive years (2022–2025). He also received the distinction of Outstanding Reviewer for IEEE AWPL for three consecutive years (2023–2025) and was named an Outstanding Associate Editor in 2025. He was selected as a 2024 IEEE AP-S Young Professional Ambassador and has been serving on the IEEE AP-S Young Professional Committee since 2025. Since late 2023, he has been a member of the IEEE AP-S Education Committee, and he is currently chairing the IEEE AP-S Student Paper Competition Committee. He also currently serves on the EurAAP Working Groups and the IEEE AP-S Technical Committee 4 on Metamaterials.

\end{IEEEbiography}

\begin{IEEEbiography}[{\includegraphics[width=1in,height=1.25in,clip,keepaspectratio]{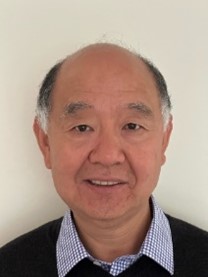}}]{Y. Jay Guo}
(Life Fellow, IEEE) received a Bachelor’s Degree and a Master’s Degree from Xidian University in 1982 and 1984, respectively, and a Ph.D. Degree from Xi'an Jiaotong University in 1987, all in China. His current research interests include 6G antennas, mm-wave and THz communications and sensing systems as well as big data technologies. He has published six books and over 700 research papers, and he holds 27 international patents.

Jay is a Fellow of the Australian Academy of Engineering and Technology, Royal Society of New South Wales and IEEE. He has won a number of the most prestigious Australian national awards including the Engineering Excellence Awards (2007, 2012) and CSIRO Chairman’s Medal (2007, 2012). He was named one of the most influential engineers in Australia in 2014 and 2015, and Australia’s Research Field Leader in Electromagnetism by the Australian Research Awards for four consecutive years since 2020. Together with his students and postdocs, he has won numerous best paper awards. In 2023, Jay received the prestigious IEEE APS Sergei A. Schelkunoff Transactions Paper Prize Award.

Jay is a Distinguished Professor and the Director of Global Big Data Technologies Centre (GBDTC) at the University of Technology Sydney (UTS), Australia. He is the founding Technical Director of the New South Wales (NSW) Connectivity Innovation Network (CIN) funded by NSW Telco Authority. He is also the Founding Director of the TPG-UTS Network Sensing Lab funded by TPG Telecom. Before joining UTS in 2014, he served as a Director in CSIRO for over nine years. Prior to CSIRO, he held various senior technology leadership positions in Fujitsu, Siemens and NEC in the U.K. 
\end{IEEEbiography}

\end{document}